\immediate\write18{bibtex \jobname}  

\documentclass[aip,apl,reprint]{revtex4-2}

\usepackage{colortbl}
\usepackage{xspace}
\usepackage{graphicx} 
\usepackage{dcolumn} 
\usepackage[dvipsnames]{xcolor}

\definecolor{lightgray}{gray}{0.9}
\definecolor{myyel}{cmyk}{0,0,0.5,0}
\definecolor{myred}{RGB}{255,180,180}
\definecolor{white}{rgb}{1,1,1}
\usepackage{color}
\usepackage[utf8]{inputenc}
\usepackage[T1]{fontenc} 
\usepackage{mathptmx}
\usepackage{newtxtext}
\usepackage{comment}
\usepackage{psfrag}
\usepackage{mathrsfs}
\usepackage{latexsym}
\usepackage{amsmath,amssymb,amsthm,mathtools}
\usepackage{wasysym}
\usepackage[export]{adjustbox}
\usepackage{xpatch}
\usepackage{multirow}

\xapptocmd\normalsize{%
 \intextsep=9pt plus 3pt minus 6pt
 \abovecaptionskip=9pt plus 3pt minus 9pt
 \belowcaptionskip=0pt plus 3pt minus 3pt
 \abovedisplayskip=9pt plus 3pt minus 9pt 
 \abovedisplayshortskip=0pt plus 3pt
 \belowdisplayskip=9pt plus 3pt minus 9pt
 \belowdisplayshortskip=5pt plus 3pt minus 4pt
}{}{}

\usepackage[activate={true,nocompatibility},final,tracking=true,kerning=true,spacing=true,factor=1100,stretch=10,shrink=10]{microtype}
\microtypecontext{spacing=nonfrench} 

\graphicspath{
{./}
}

\newcommand{\delI}{\ensuremath{\delta I}\xspace}

\newcommand{\Lmint}{\ensuremath{L_{\rm m}}\xspace}
\newcommand{\Lmext}{\ensuremath{L_{\rm ext}}\xspace}

\newcommand{\mlsi}{\texttt{3D-MLSI}\xspace}
\newcommand{\FH}{\texttt{FastHenry}\xspace}
\newcommand{\induct}{\texttt{induct}\xspace}

\newcommand{\Lo}{\ensuremath{L_{\rm 0}}\xspace}
\newcommand{\Lk}{\ensuremath{L_{\rm kin}}\xspace}
\newcommand{\Jc}{\ensuremath{J_{\rm c}}\xspace}
\newcommand{\Ic}{\ensuremath{I_{\rm c}}\xspace}
\newcommand{\betal}{\ensuremath{\beta_L}\xspace}
\newcommand{\Lsq}{\ensuremath{L_{\rm sq}}\xspace}
\newcommand{\Tc}{\ensuremath{T_{\rm c}}\xspace}
\newcommand{\LCO}{\ensuremath{L_{\rm co}}\xspace}
\newcommand{\LPAR}{\ensuremath{L_{\rm par}}\xspace}

\newcommand{\YBCO}{YBa$_2$Cu$_3$O$_{7-x}$\xspace}
\newcommand{\lzero}{\ensuremath{\ell_0}\xspace}
\newcommand{\lone}{\ensuremath{\ell_1}\xspace}
\newcommand{\cmmnt}[1]{\ignorespaces}

\newcommand{\Lw}{\ensuremath{L_{\rm w}}\xspace}

\newcommand{\Aeffsq}{\ensuremath{A_{\rm eff}^{\rm sq}}\xspace}
\newcommand{\Aeffw}{\ensuremath{A_{\rm eff}^{\rm w}}\xspace}
\newcommand{\Aeffm}{\ensuremath{A_{\rm eff}^{\rm m}}\xspace}
\newcommand{\Bz}{\ensuremath{B_z}\xspace}
\newcommand\xyplane{$x$\nobreakdash--$y$~plane\xspace}
\newcommand{\Mf}{\ensuremath{M_{\rm f}}\xspace}

\newcommand{\rb}[1]{\raisebox{1.5ex}[0pt]{#1}}
\newcolumntype{d}[1]{D{.}{.}{#1}}

\makeatletter
\def\CT@@do@color{%
	\global\let\CT@do@color\relax
	\@tempdima\wd\z@
	\advance\@tempdima\@tempdimb
	\advance\@tempdima\@tempdimc
	\advance\@tempdimb\tabcolsep
	\advance\@tempdimc\tabcolsep
	\advance\@tempdima2\tabcolsep
	\kern-\@tempdimb
	\leaders\vrule
	\hskip\@tempdima\@plus  1fill
	\kern-\@tempdimc
	\hskip-\wd\z@ \@plus -1fill }
\makeatother

\begin{document}

\title[]{Determining the temperature-dependent London penetration depth in HTS thin films and its effect on SQUID performance}

\author{Shane Keenan}
\thanks{Authors to whom correspondence should be addressed: \\shane.keenan@csiro.au, colin.pegrum@strath.ac.uk}
\affiliation{CSIRO Manufacturing, West Lindfield, NSW 2070, Australia}

\author{Colin Pegrum}
\affiliation{Department of Physics, University of Strathclyde, Glasgow G4 0NG, UK}

\author{Marc Gali Labarias}
 \affiliation{CSIRO Manufacturing, West Lindfield, NSW 2070, Australia}

\author{Emma E Mitchell}
 \affiliation{CSIRO Manufacturing, West Lindfield, NSW 2070, Australia}
 


\begin{abstract}

The optimum design of high-sensitivity Superconducting Quantum Interference Devices (SQUIDs) and other devices based on thin HTS films requires accurate inductance modeling. This needs the London penetration depth $\lambda$ to be well defined, not only at 77\,K, but also for any operating temperature, given the increasingly widespread use of miniature low-noise single-stage cryocoolers. Temperature significantly affects all inductances in any active superconducting device and cooling below 77\,K can greatly improve device performance, however accurate data for the temperature dependence of inductance and $\lambda(T)$ for HTS devices is largely missing in the literature. We report here inductance measurements on a set of 20 different thin-film \YBCO SQUIDs at 77\,K with thickness $t = 220$ or 113 nm. By combining experimental data and inductance modeling we find an average penetration depth $\lambda(77) = 391\,$nm at 77\,K, which was independent of $t$. Using the same methods we derive an empirical expression for $\lambda(T)$ for a further three SQUIDs measured on a cryocooler from 50 to 79\,K. Our measured value of $\lambda(77)$ and our inductance extraction procedures were then used to estimate the inductances and the effective areas of directly coupled SQUID magnetometers with large washer-style pick-up loops. The latter agree to better than 7\% with experimentally-measured values, validating our measured value of $\lambda(77)$ and our inductance extraction methods.

\end{abstract}


\maketitle


SQUIDs are the most sensitive magnetic field detectors with a wide range of proven applications outside the laboratory, \cite{keenan:298,Fagaly_Wiley_EncEEE,stolz:033001} such as airborne,\cite{lee:468} ground\cite{leslie:70} and marine \cite{keenan:025029} surveying, archaeology\cite{linzen:71} or non-destructive evaluation.\cite{NDE_SQUID_HB} High-temperature superconductor (HTS) SQUIDs made from \YBCO (YBCO) are especially attractive for such uses, as they can be used in liquid nitrogen at 77\,K and also in compact single-stage cryocoolers, where cooling below 77\,K can enhance their performance significantly.

For thin-film SQUIDs and related devices the inductances of all their parts need to be known at the design stage in order to optimize performance at a given temperature $T$. These inductances depend significantly on $T$ due to the changing penetration depth $\lambda(T)$, especially approaching the transition temperature \Tc, as our data reported here will show. There are very few recent measurements of $\lambda$ similar to ours on high-quality YBCO films. This motivated us to make direct inductance measurements on a set of YBCO SQUIDs with different line-widths, loop sizes and film thicknesses and to combine these with inductance extraction techniques to find the penetration depth $\lambda(77)$ at 77\,K and also an expression for $\lambda(T)$ in the range $50 < T < 78$\,K. 

Design can be supported by various inductance extraction tools or formulas available for self inductance, but whatever the method, these all need $\lambda(T)$ as a parameter. For example, the inductance \Lsq of a SQUID must be well known to optimize the modulation parameter \betal = $2\Lsq\Ic/\Phi_0$ (\Ic is the junction critical current and $\Phi_0$ is the flux quantum). This is also true for the loops in Superconducting Quantum Interference Filters (SQIFs) \cite{mitchell:06LT01}. SQIFs have a large dynamic range \cite{oppenlander:936} that makes them well-suited to operation below 77\,K in a cryocooler and their loop inductances need to be known for the chosen operating temperature. Inductances of the component parts of direct-coupled SQUID magnetometers \cite{lam:123905} also need to be known accurately to determine their effective area.

\begin{figure}[h!]
\includegraphics[width=0.38\columnwidth,valign=c]{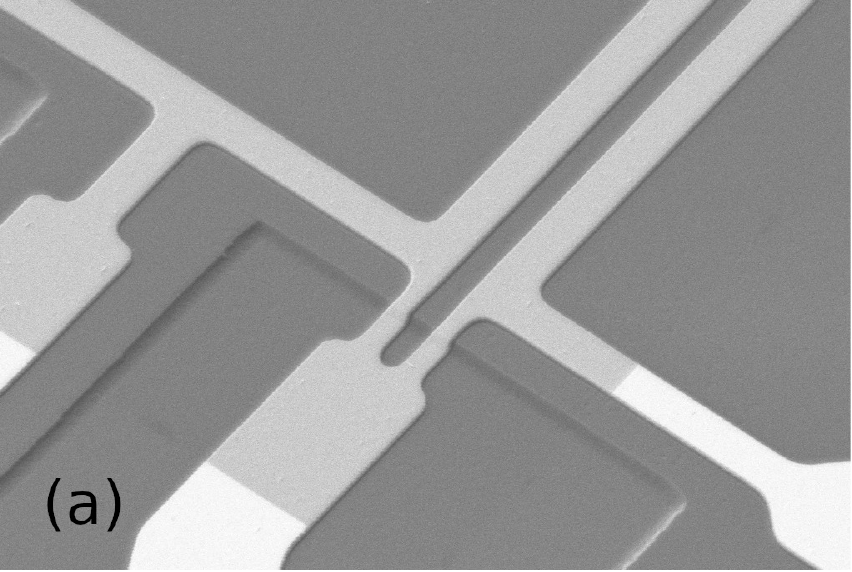}
\hspace{0.05\columnwidth} 
\includegraphics[width=0.49\columnwidth,valign=c]{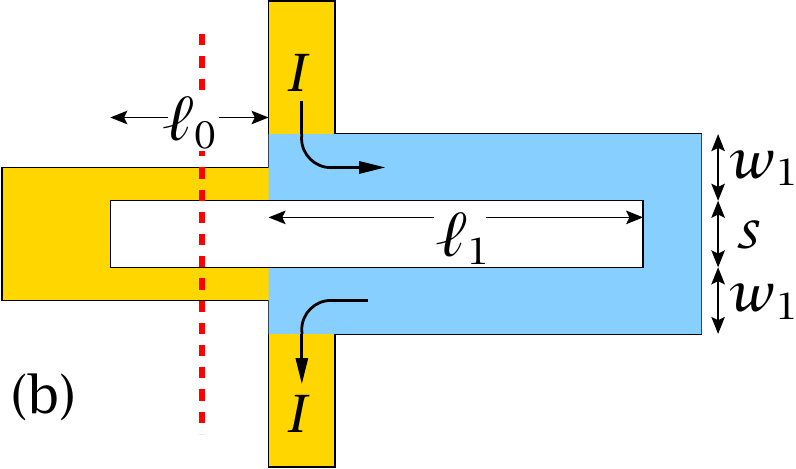}
\caption{\label{fig:sq_design} (a) SEM  of part of a SQUID with step-edge junctions $2\,\mu$m wide. (b) The SQUID model. Injected current $I$ flows through the CPS part, shown in blue. The red dashed line indicates the step edge.}
\end{figure}

One way to measure $\lambda$ in HTS thin films is the two-coil mutual inductance technique first developed by Fiory et al.\ \cite{fiory:2165} or variations on this.\cite{claassen:3028,wang:3865,he:113903} More advanced techniques include muon spin rotation and microwave resonance.\cite{prozorov:R41,he:113903} All these are limited to large millimeter-sized samples. Our approach is to  deduce $\lambda(T)$ from direct measurements of inductance in SQUIDs, like those in Fig.~\ref{fig:sq_design}.

These all have a co-planar strip-line (CPS) design, shorted at one end, with $2\,\mu$m wide step-edge junctions. This is the design used in our directly-coupled magnetometers \cite{lam:123905} which we discuss later. We measure  the coupled self inductance \LCO of the part of the SQUID loop shown in blue in Fig.~\ref{fig:sq_design}(b), which has length \lone. An external current $I$ is fed through this (in our magnetometers this would be the current sensed by the directly-coupled pick-up loop, which is absent for these measurements). The total SQUID inductance $\Lsq = \LCO + \LPAR$ where \LPAR is the parasitic inductance of the uncoupled part with length \lzero. From the change \delI causing an output change of one flux quantum $\Phi_0$, $\LCO=\Phi_0/\delI$. A technique like this was used for HTS SQUIDs by Forrester et al.\ \cite{forrester:1835} and subsequently by others. \cite{grundler:5273,fuke:L1582,mitchell:1282,li:1600404,ruffieux:025007,ruffieux:asc2020,shimazu:1451}

Measurements similar to ours have been made by several others on SQUIDs to find $\lambda$. Li et al. \cite{li:1600404} measured \Lsq for five nano-slit SQUIDs fabricated by a focused helium ion beam on a 25\,nm thick YBCO film and found $\lambda = 180$\,nm gave the best match to experiment at 9\,K.  Ruffieux et al.\ \cite{ruffieux:025007,ruffieux:asc2020} measured the inductance of ten HTS SQUIDs fabricated from two separate 140\,nm thick YBCO films on SrTiO$_3$ substrates, one with a CeO$_2$ buffer layer and one without, with varying \lone. %
They reported $\lambda$ in the ranges $542-648$\,nm for no CeO$_2$ and $313-426$\,nm with CeO$_2$. They saw large differences in $\lambda$ between SQUIDs on the same sample, which they attributed to differences in film thickness and \Tc across the sample and slight temperature variation during measurements. Mitchell et al. \cite{mitchell:1282} made earlier inductance measurements on seven different CPS SQUIDs made from 300\,nm YBCO films and found the inductance per unit length $L_0=0.97\,{\rm pH}/\mu$m for $w_1=2\,\mu$m and $s=4\,\mu$m.

Our present work was done in three related parts. Firstly, we measured \LCO for twenty bare SQUIDs of different geometries at 77\,K (with no magnetometer loop attached) and used inductance extraction models to determine $\lambda(77)$. The SQUID properties are listed in Table~\ref{tab:squid_props}; they are in five sets with different properties and for each set \lone has four different values. For each set we measured $\LCO(\lone)$ and derived the inductance per unit length $L_0$ of the CPS by linear regression.  We combined our experimental measurements of $L_0$ with inductance extraction data from SQUID models in which $\lambda$ is a variable parameter, to find $\lambda(77)$ for the five different samples.

In the second part we measured \LCO(T) in a cryocooler for three different SQUIDs on one chip, to extract an empirical expression for $\lambda(T)$. We measured $\LCO(T)$ for each, for $50 < T < 78$\,K.  By inductance extraction we found an expression for $\LCO(\lambda)$, merged this with the experimental $\LCO(T)$ and fitted this data to get an expression for $\lambda(T)$.

In the concluding part we used our value of $\lambda(77)$ and our inductance extraction techniques to estimate the effective area of washer-coupled SQUID magnetometers and compared these with measured effective areas.
 

\begin{table}[t]
	\caption{\label{tab:squid_props}Summary of the film thickness $t$, \Tc and dimensions of twenty SQUIDs in five sets used in the first part of this study. They are on three separate substrates S1, S2 and S3 and from three different film batches B1, B2 and B3.  Each set has four different values of $\ell_1$. All have the layout shown in Fig.~\ref{fig:sq_design} with $\lzero=10\,\mu$m.}
	\begin{ruledtabular}
		\begin{tabular}{l@{\hspace{-2ex}}r@{\hspace{-2ex}}lllll@{\hspace{-2ex}}l@{}}
			SQUID set        && S1-A & S2-A & S3-A & S3-B & S3-C &\\
			Batch            && B1 & B2 & B3 & B3 & B3 &\\
			$T_{\rm c}$ (K)  && 85.9 & 86.2 & 87.5 & 87.5 & 87.5 &\\
			$t$ (nm)         && 220  & 113  & 113  & 113 & 113 &\\
			$s$ ($\mu$m)     && 4 &  4 & 4  & 4    & 2 &\\
			$w_1$ ($\mu$m)   && 4 &  4 & 4  & 8    & 8 &\\
			\multirow{2}{*}{$\ell_1$ ($\mu$m)} &\multirow{2}{*}{$\Big\{$}&  50,\,75, & 20,\,34, & 20,\,34, & 20,\,30, & 20,\,30,&\multirow{2}{*}{$\Big\}$}\\
			&& 100,\,125 & 44,\,50  & 44,\,50  & 40,\,50  & 40,\,50 &\\
		\end{tabular}
	\end{ruledtabular}
\end{table}


\LCO depends on device structure and film properties and has three contributions, from internal and external magnetic energy,  \Lmint and \Lmext respectively, plus a kinetic contribution \Lk.  In general $ \Lk\gg \Lmint $. Both \Lmint and \Lk depend on $\lambda$; as $\lambda $ increases \Lk diverges rapidly, whereas \Lmext reaches a limiting value as field penetration nears completion.  For a single isolated line \Lk is approximately $(\mu_0\lambda^2)/wt$ per unit length,\cite{lee:2419} where $\mu_0$ is the magnetic permeability and $w$ and $t$ are its width and thickness. For the lines in the CPS we use, especially ones with a narrow spacing $s$, \Lk differs somewhat from this expression.\cite{yoshida:3844} We chose to determine inductances using the extraction package \mlsi,\cite{khapaev:1090,khapaev:055013} rather than from formulas which may have limited accuracy for our designs. \mlsi is a Finite Element Method (FEM) and can handle thin-film structures of any arbitrary geometry. It extracts the total inductance $ \Lk+\Lmint+\Lmext $.

Our SQUIDs were fabricated from 113 and 220\,nm of YBCO on one side of polished 1\,cm$^2$ MgO substrates using step-edge junctions. \cite{foley:4281,mitchell:065007} The films were deposited by Ceramic Coating GmbH \cite{ceraco} by reactive co-evaporation, who measured the \Tc values listed in Table~\ref{tab:squid_props} inductively\cite{fiory:2165,claassen:3028} on unpatterned samples. We have also made resistance-temperature measurements on 4\,$\mu$m tracks within patterned devices and find post-deposition processing causes negligible reduction in \Tc. All films of both thicknesses have a similar current density, $\Jc = 3.1-3.2\,{\rm MA/cm}^2$. Devices were patterned and etched by our standard photo-lithography procedure.\cite{foley:4281}  We have reported previously current-voltage characteristics of our junctions at 77\,K \cite{mitchell:065007} and typical voltage-flux responses for SQUIDs using them.\cite{lam:123905}

In the first part of our work we measured \LCO for the twenty SQUIDs in liquid nitrogen inside six concentric mu-metal shields, using a STAR Cryoelectronics PCI-1000 control unit.\cite{star} The encapsulated chip was mounted on a radio-frequency shielded probe. Each SQUID was current biased to its optimal operating point (maximum peak to peak voltage modulation) and flux varied by the current $I$. The PCI-1000’s 12-bit high-resolution internal current generator supplied $I$, which was averaged over a minimum of 10 $\Phi_0$ to improve accuracy. We used a 5\,Hz triangle waveform for this and a transfer coefficient of 10\,$\mu$A/V. The amplitude was then varied to couple $10\,\Phi_0$ or more into the SQUID. The PCI-1000 user manual has full details. \cite{star}

\begin{figure}[h!]
\centering
    \includegraphics[height=0.9\columnwidth,angle=-90]{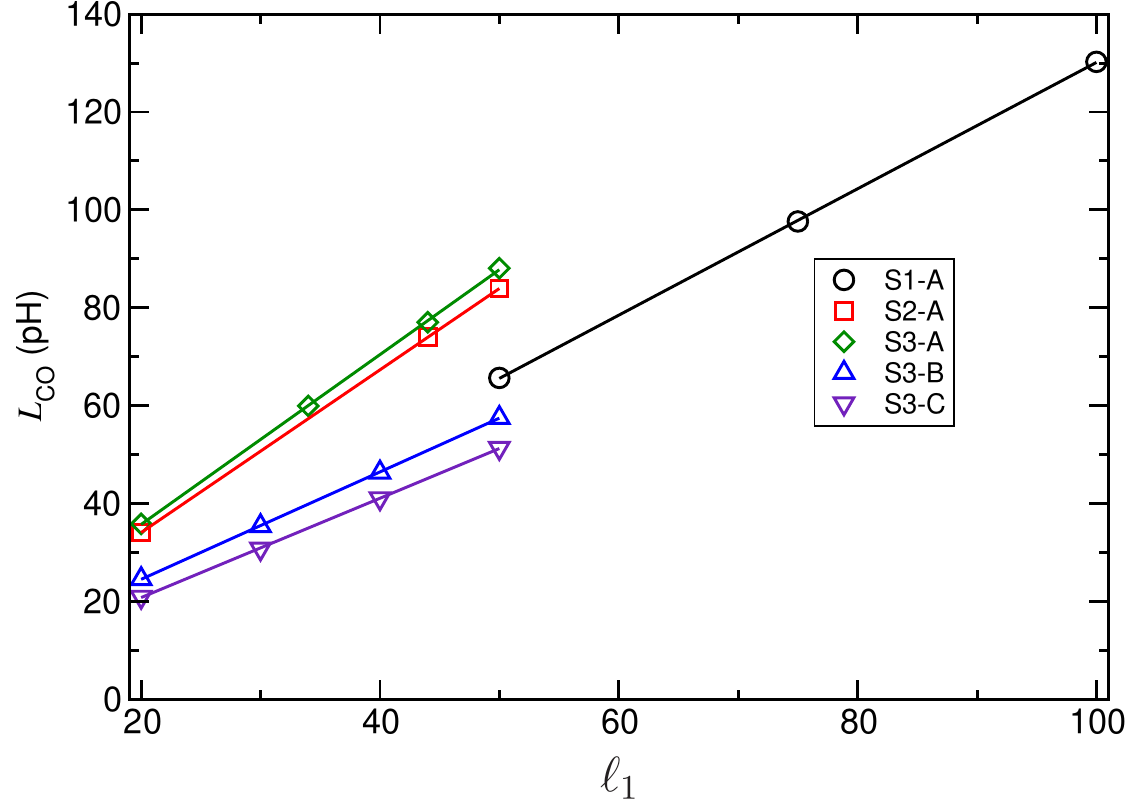} 
    \caption{\label{fig:meas_LCO} The measured coupled SQUID inductance \LCO vs its length \lone for the five different SQUID sets. Each solid line is a linear fit to the data points and its gradient is \Lo.}
\end{figure}

Fig.~\ref{fig:meas_LCO} plots the experimental \LCO as a function of \lone for each of the five SQUID sets and shows a highly linear dependence: $\LCO=\Lo\lone +L_{\rm end}$. $L_{\rm end}$ is the same for all SQUIDs in each set and is due to small extra contributions to inductance at the shorted end and around the terminals.  Experimental values of \Lo are listed in Table~\ref{tab:L_values}, along with our estimates of $\lambda$ at 77\,K based on these data, using the following procedures. 

\begin{figure}
\centering
\includegraphics[width=0.95\columnwidth]{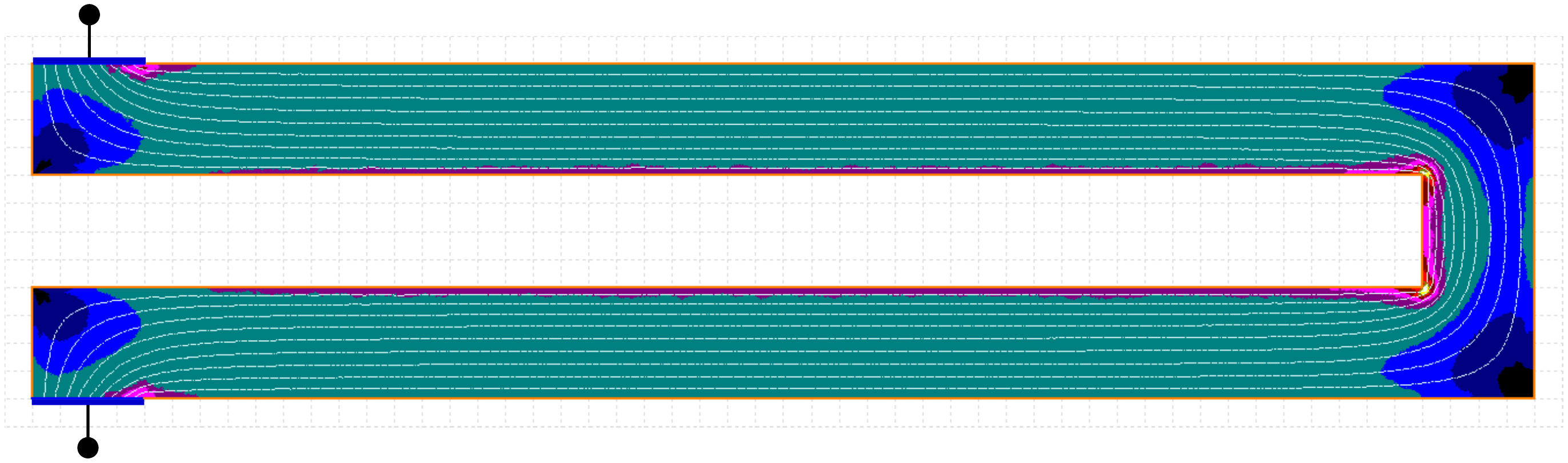} 
\includegraphics[width=0.95\columnwidth]{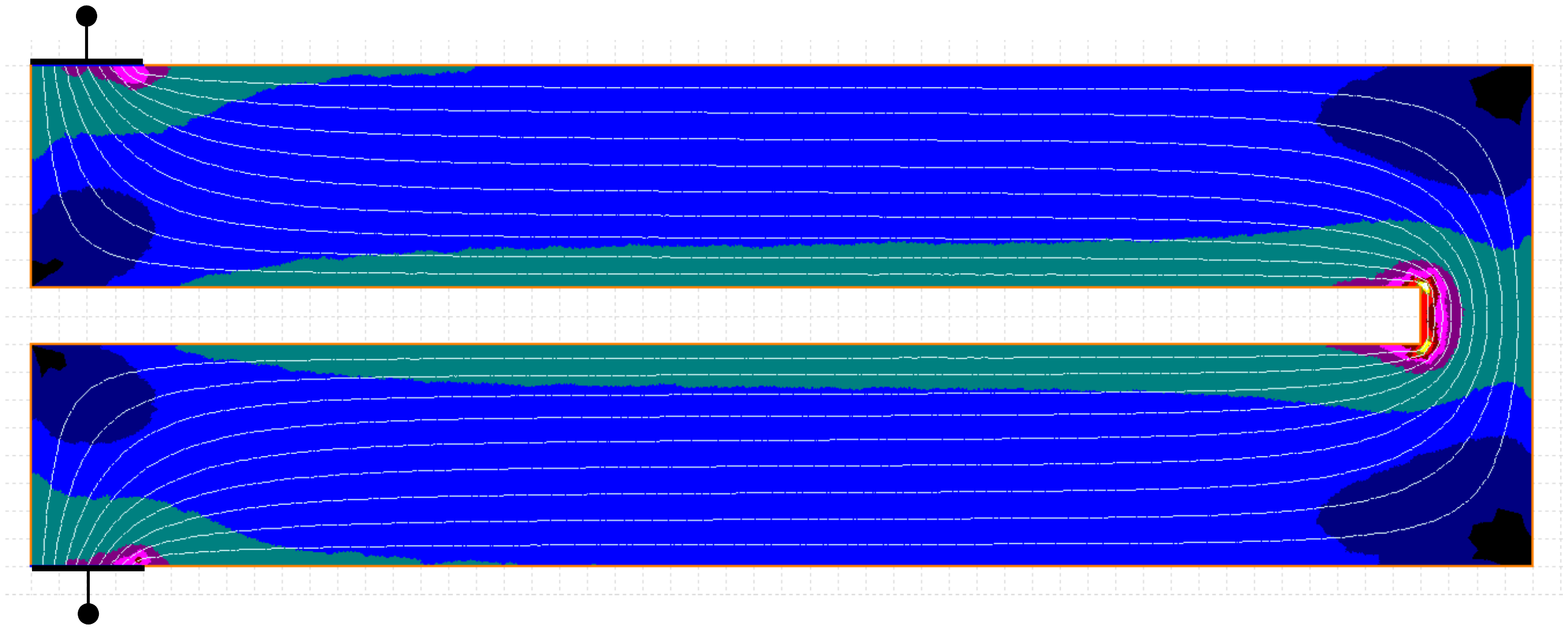} 
\caption{\label{fig:mlsi}Layouts and stream function solutions for SQUIDs S1-A (SQ1) and S3-C (SQ4), both with $\lone = 50\,\mu$m, for $\lambda=400\,$nm. Current flows in and out through $4\,\mu$m wide terminals marked {\color{Black}$\CIRCLE$}. Colors represent the magnetic field generated, black lowest, red highest, on a normalised scale of 0 to 1.0.}
\end{figure}

To estimate $\lambda(77)$ we use \mlsi to extract \Lo for each style of SQUID. \mlsi also visualizes current flow, as in Fig.~\ref{fig:mlsi}. Current bunching is evident at the closed end and around the terminals. This adds extra inductance to \LCO which, as for the experiment, we eliminated by extracting \LCO for a range of lengths \lone and regression to $\LCO(\lone$). For comparison we also used \FH \cite{kamon:1750} and the program \induct, \cite{induct} derived from work by Chang. \cite{chang:764}  \induct derives \Lo directly for infinitely-long lines. For \FH we extracted the inductance $L_{\rm CPS}$ of a pair of anti-parallel open-circuit lines for a range of lengths $\ell$. Again, regression to $L_{\rm CPS}(\ell)$ was done to find \Lo, to avoid end effects to do with the point-injection of current into lines of finite width by \FH. These three extraction tools all include \Lk. For all of them $\lambda$ is a user-defined parameter, but the form of its temperature dependence is \textit{not}. 

Each extractor was run for $100 < \lambda < 500\,$nm in 25\,nm steps.  The resulting $\Lo-\lambda$ data-sets were fitted to a polynomial expression for $\lambda(\Lo)$.  We then used the experimental value of \Lo to derive our estimate of $\lambda(77)$ for each of the five SQUID designs. Table~\ref{tab:L_values} brings together these data for all three extraction methods. We have greatest confidence in the values of $\lambda(77)$ using the FEM package\mlsi, though for extracting \Lo for the CPS structure both \FH and \induct give similar answers, as Table~\ref{tab:L_values} confirms.

\makeatletter
\newcommand{\vast}{\bBigg@{2.9}}
\makeatother

\begin{table}
	\caption{\label{tab:L_values} Measured data and the values deduced for $\lambda$.  Each of the five sets has four SQUIDs with different values of $\ell_1$. $\LCO=\Phi_0/\delI$ was measured for each SQUID and $L_0$ was derived from the $\LCO(\ell_1)$ data for each set. $\lambda_{\rm mlsi}$, $\lambda_{\rm ind}$ and $\lambda_{\rm fh}$ are from fits to inductance extraction data using \mlsi, \induct and \FH models respectively.}
		\begin{tabular}{ccd{0}ld{2}llll}
			\hline\hline
			Set & SQUID & \multicolumn{1}{c}{$\ell_1$} & $\delta I$ & \multicolumn{1}{c}{$L_{\rm CO}$} & $L_0$ & $\lambda_{\rm mlsi}$ & $\lambda_{\rm ind}$ & $\lambda_{\rm fh}$ \\
			&   &  \multicolumn{1}{l}{($\mu$m)} & ($\mu$A) & \multicolumn{1}{c}{(pH)} & (pH/$\mu$m)  & (nm) & (nm) & (nm) \\
			\hline
		    \rowcolor{myyel}& SQ1 &  50  & 31.48 &  65.69 &&&&\\
			\rowcolor{myyel}& SQ2 &  75  & 21.18 &  97.64 &&&&\\
			\rowcolor{myyel}\rb{\textbf{S1-A}} & SQ3    &  100 & 15.88 & 130.23 & \rb{1.291} & \rb{402}& \rb{401} & \rb{404}\\  
			\rowcolor{myyel}& SQ4     &  125 & $-$   &    -   &&&&\\
			&&&&&&&&\\[-1ex]
			\rowcolor{myred}& SQ1 &  20  & 60.66 & 34.09  &&&&\\ 
			\rowcolor{myred}& SQ2      &  34  & $-$   &  -     &&&&\\
			\rowcolor{myred}\rb{\textbf{S2-A}} &SQ3      &  44  & 27.94 & 74.01  & \rb{1.669} &  \rb{386} & \rb{385} & \rb{387} \\
			\rowcolor{myred}& SQ4      &  50  & 24.64 & 83.92  &&&&\\
			&&&&&&&&\\[-1ex]
			\rowcolor{myyel}& SQ1 &  20  & 57.66 & 35.86  &&&&\\  
			\rowcolor{myyel}& SQ2      &  34  & 34.48 & 59.97  &&&&\\   
			\rowcolor{myyel}\rb{\textbf{S3-A}} & SQ3      &  44  & 26.84 & 77.04  &\rb{1.734} & \rb{401} & \rb{399} & \rb{401}\\
			\rowcolor{myyel}& SQ4      &  50  & 23.48 & 88.07  &&&& \\
			&&&&&&&&\\[-1ex]
			\rowcolor{myred}&  SQ1 &  20  & 84.00 & 24.62  &&&&\\ 
			\rowcolor{myred}& SQ2      &  30  & 58.34 & 35.44  &&&&\\    
			\rowcolor{myred}\rb{S3-B} & SQ3      &  40  & 44.58 & 46.38  & \rb{1.097} & \rb{384} & \rb{370} & \rb{377} \\  
			\rowcolor{myred}& SQ4      &  50  & 35.94 & 57.54  &&&&\\
			&&&&&&&&\\[-1ex]
			\rowcolor{myyel}& SQ1 &  20  & 98.68 & 20.95  &&&&\\
			\rowcolor{myyel}& SQ2      &  30  & 67.16 & 30.79  &&&&\\  
			\rowcolor{myyel}\rb{\textbf{S3-C}} & SQ3      &  40  & 50.36 & 41.06  & \rb{0.984} & \rb{385} & \rb{359} & \rb{363} \\ 
			\rowcolor{myyel}& SQ4      &  50  & 40.28 & 51.34  &&&&\\
			\hline\hline
		\end{tabular}
\end{table}

\begin{figure}[h!]
\centering
\includegraphics[width=0.49\columnwidth]{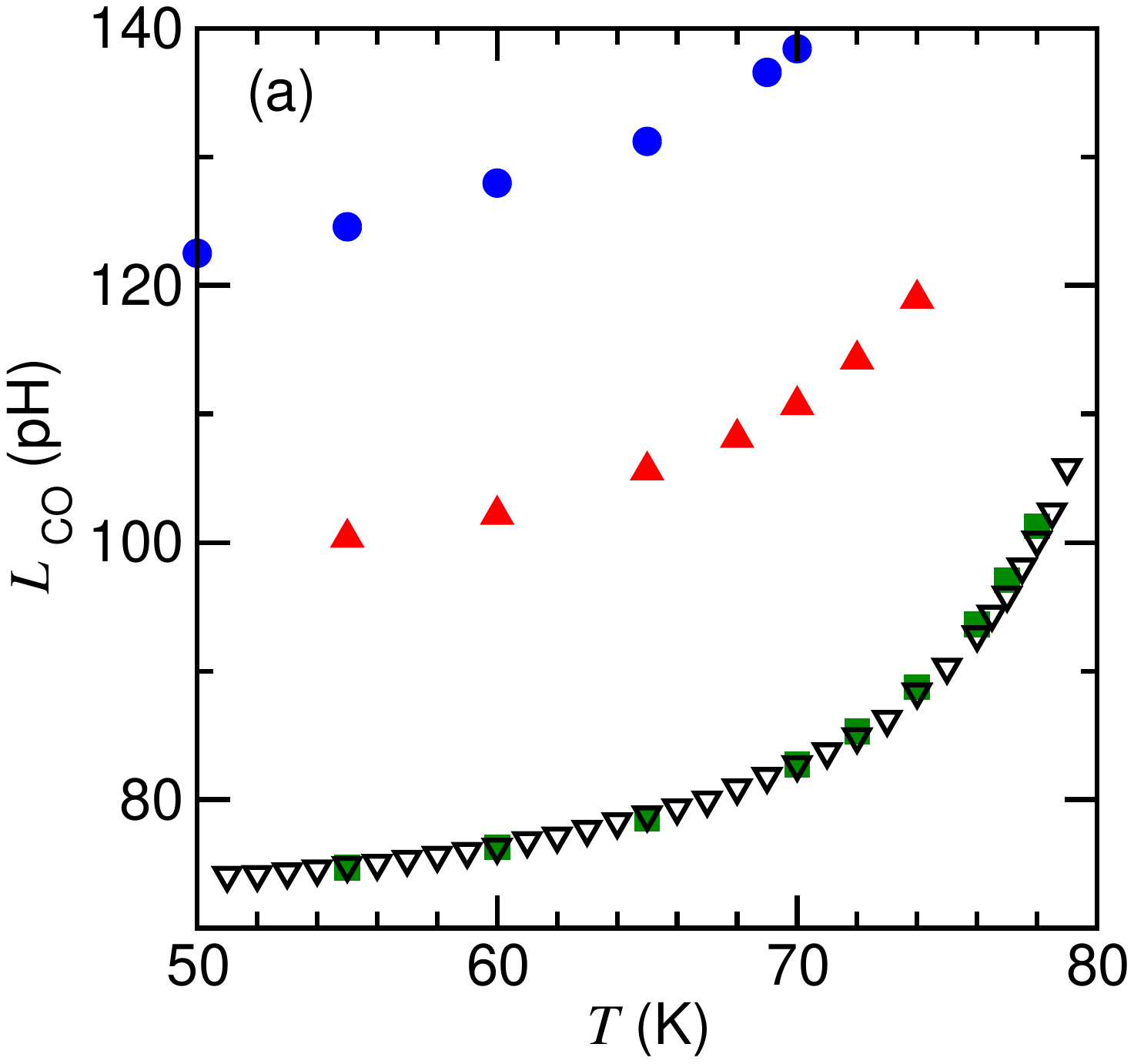} 
\includegraphics[width=0.49\columnwidth]{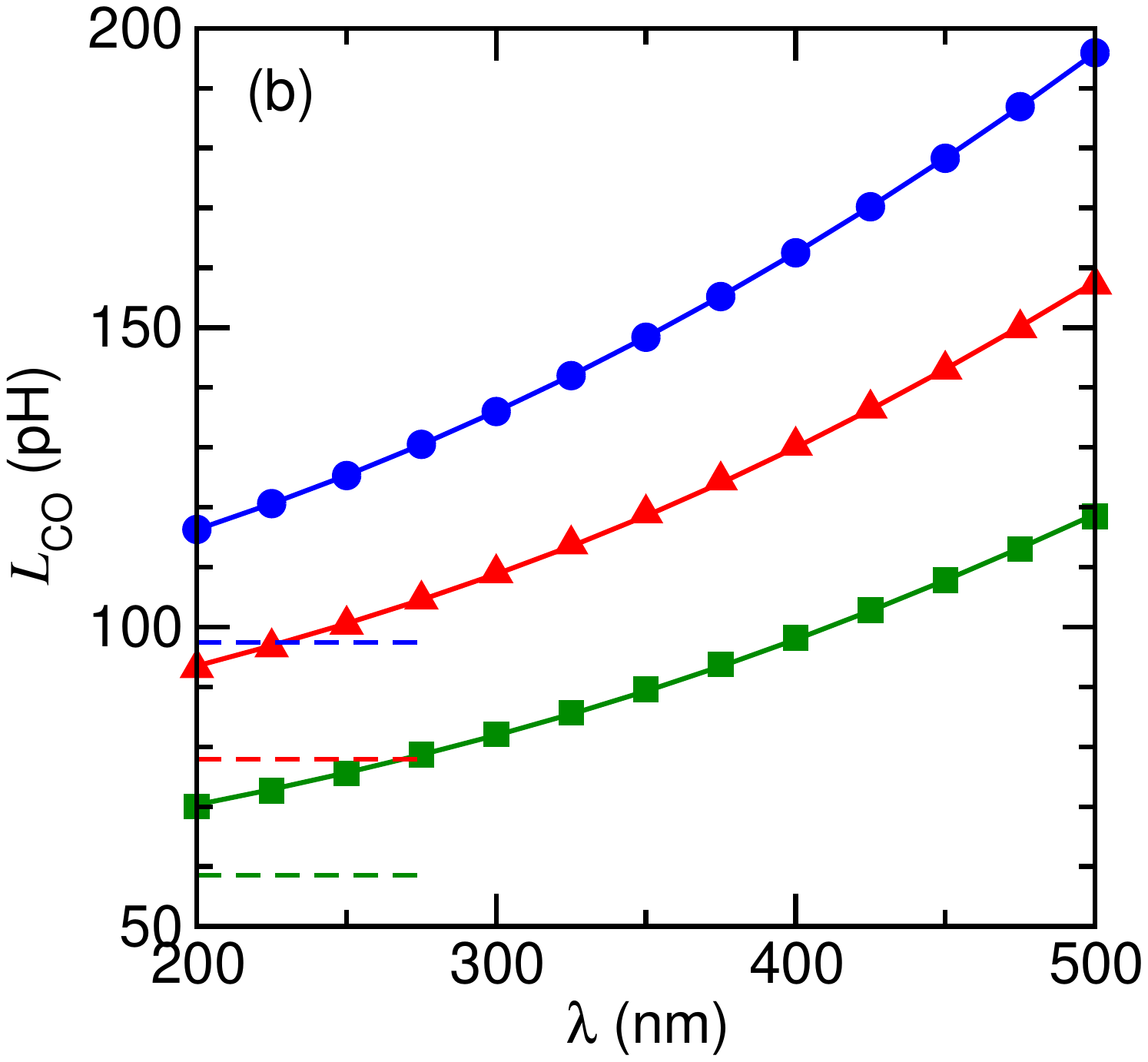} 
\caption{\label{fig:L_lambda_T}(a) Measured inductance \LCO as a function of temperature $T$ for three SQUIDs from set S1-A {\color{Blue}$\CIRCLE$}~SQ4; {\color{Red}$\blacktriangle$}~SQ3; $\triangledown$ and {\color{Green}$\blacksquare$}~SQ2. Devices SQ4 and SQ3 had smaller flux modulation, owing to their higher \Lsq  which limited measurements of \LCO to less than 70 and 74\,K respectively. The two data-sets for SQ2 were recorded on different occasions. (b) \LCO extracted by \mlsi, varying $\lambda$, for the same set of SQUIDs. Solid lines are cubic fits. Dashed lines indicate values of \Lmext for each SQUID.}
\end{figure}

In the second stage of our work we used a cryocooler to measure three SQUIDs, one by one, from set S1-A, to get $\LCO(T)$ for each, as in Fig.~\ref{fig:L_lambda_T}(a).  The temperature was measured by a PT-111 RTD sensor, mounted on the copper block carrying the substrate, 6\,mm from its edge. A Lakeshore 335 controller with a $50\,\Omega$ heater regulated the temperature to $\approx\pm 0.1$\,K.

\begin{figure}[h]
\centering
\includegraphics[width=0.8\columnwidth]{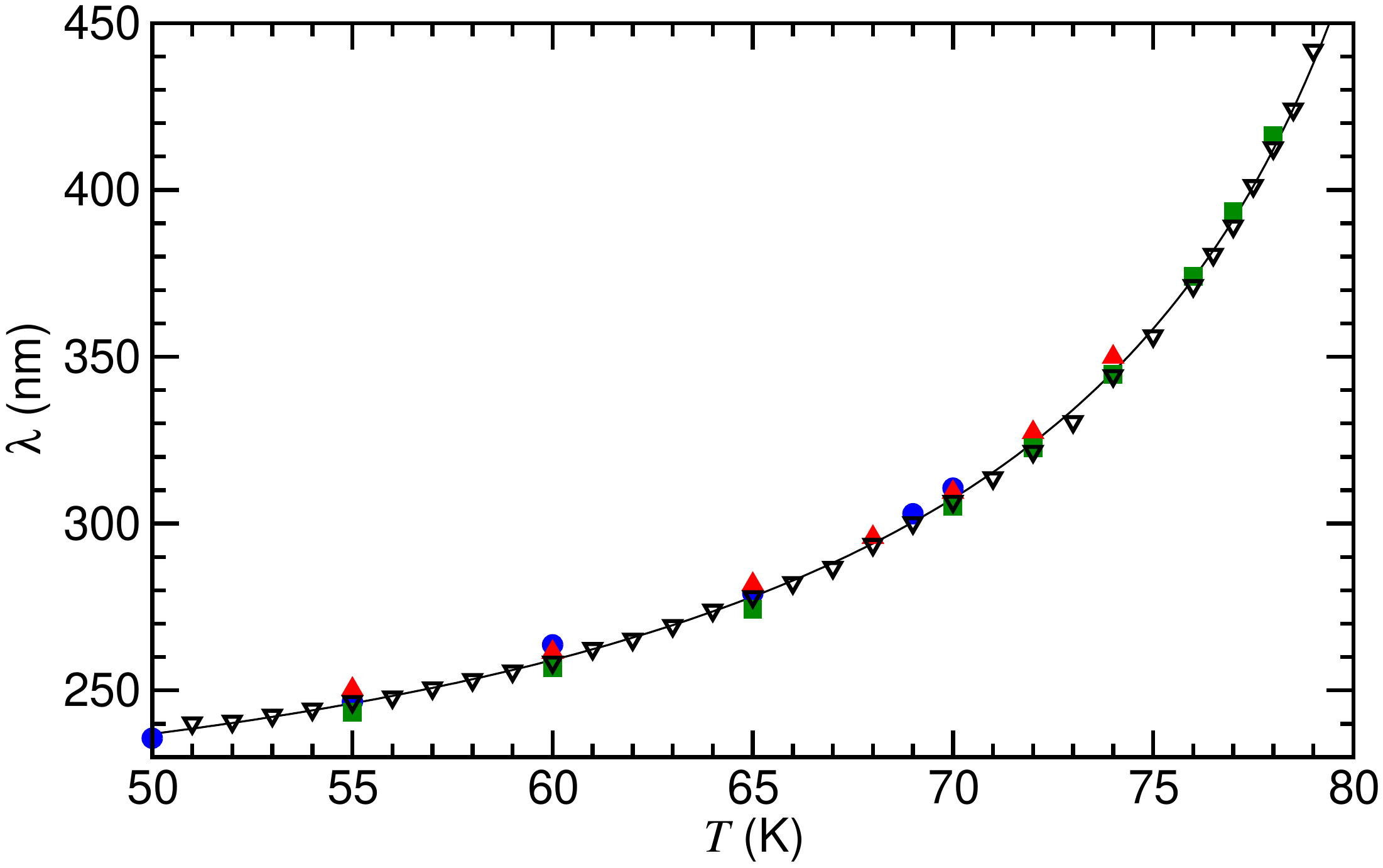}
\caption{\label{fig:L_T_all_fit} $\lambda(T)$ derived from the data shown in Figs.~\ref{fig:L_lambda_T}(a) and (b) for the same three SQUIDs.  Data markers are the same as those in Fig.~\ref{fig:L_lambda_T}. The solid line is a fit to (\ref{eq:lambda_T}) with $\Tc= 85.9\,$K.}
\end{figure}

 We then used \mlsi with structures like those in Fig.~\ref{fig:mlsi} to extract \LCO, for $200 < \lambda < 500\,$nm and these data were fitted to functions for $\LCO(\lambda)$, as in Fig.~\ref{fig:L_lambda_T}(b).  We then merged $\LCO(\lambda)$ with the experimental values of $\LCO(T)$ for the three SQUIDs to generate our consolidated result for $\lambda(T)$, shown in Fig.~\ref{fig:L_T_all_fit}. It is clear that our data derived from four separate measurements on three different SQUIDs have the same temperature dependence, which we believe validates our measurement of \LCO(T) and our analysis using \mlsi. This is the only extraction package that can be used, as it will correctly include the extra temperature-dependent inductance associated with current flow near the terminals and around the shorted end, as is evident in Fig.~\ref{fig:mlsi}. Neither \FH, nor \induct nor CPS formulas can include these effects. The dashed lines in  Fig.~\ref{fig:L_lambda_T}(b) are \Lmext values for each SQUID, found from \mlsi by setting $\lambda = 0$ which makes $\Lk=\Lmint=0$. This clearly shows the significant contribution from \Lk to the total inductance in this temperature range.

We made a non-linear least-squares fit to these data, assuming $\lambda(T)$ has this general form: 
\begin{equation}\label{eq:lambda_T}
 \lambda(T)=\lambda(0)\left[ 1 - \left({T}/{T_{\rm c}}\right)^P\right]^{-n}
 \end{equation}
The solid line in Fig.~\ref{fig:L_T_all_fit} has $\lambda(0)$ and $P$ as fitting parameters, with fixed $n=0.5$ and $\Tc=85.9\,$K, the measured value. This gives $\lambda(0)=217\,$nm and $P=3.36$.

The form of (\ref{eq:lambda_T}) is based on the two-fluid Gorter-Casimir ad hoc expression for the super-electron density $n_{\rm s}$ and the London expression for $\lambda(n_{\rm s})$, which give $P=4$ and $n=0.5$ for low-temperature superconductors.  There is no underlying microscopic reason to expect specific values of $P$ or indeed $n$ for our HTS thin films. Our value of $P$ exceeds the 1.94 to 2.45 range found some while ago by Il'ichev et al. \cite{ilichev:708} Others generally found $P\approx2$.\cite{forrester:1835,chen:9,brorson:6185,li:1600404,prohammer:5370} Our value of 217\,nm for $\lambda(0)$ is within the accepted range of other measurements, for example \cite{zaitsev:335,chen:9,farber:515} for transport in the $a$-$b$ plane of high-quality YBCO thin films.

\begin{figure}[h!]
\centering
\includegraphics[width=0.95\columnwidth]{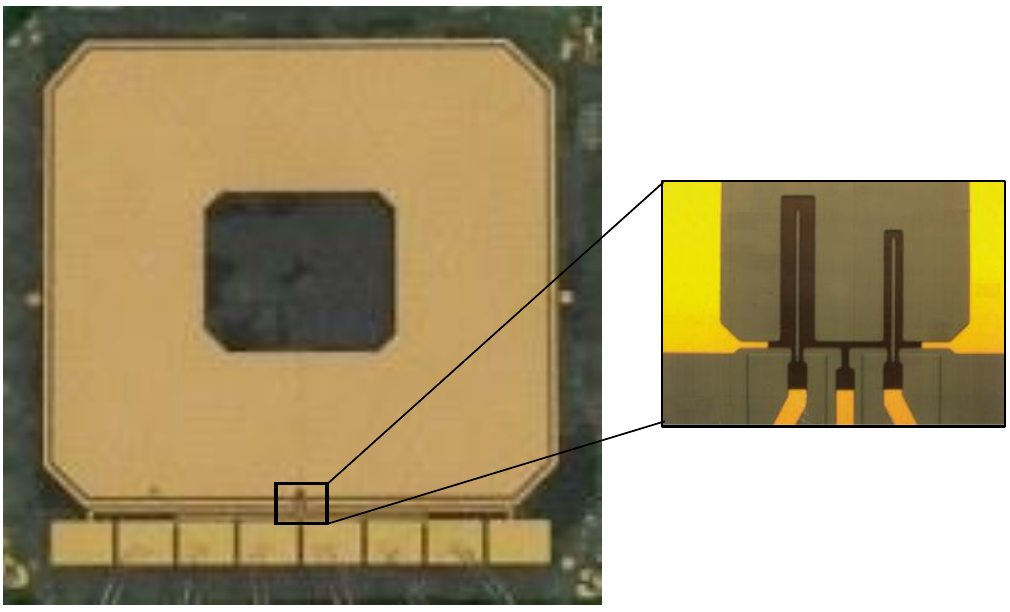} 
\caption{\label{fig:mw}A directly-coupled magnetometer with a washer-style pick-up loop on a 1\,cm$^2$ substrate.  The overall size of the washer is $8.2 \times 7.5$\,mm, the hole is $3.2 \times 2.7$\,mm and the slit in the washer is $6\,\mu$m wide. Each DCM contains two SQUIDs with \LCO in series, as shown in the inset, types D and E in this example.}
\end{figure}

To conclude our work we extended our inductance extraction methods to compare the measured and simulated effective areas of a set of our directly-coupled magnetometers (DCM's). Fig.~\ref{fig:mw} shows one of the DCM's.
 
All the DCM's have the same size of split-washer pick-up loop, directly coupled to two SQUIDs in series that can be biased individually. These SQUIDs have the same geometry as those in Fig.~\ref{fig:sq_design} and Table~\ref{tab:squid_props} but have different dimensions and are identified in Table~\ref{tab:sq_results} as types D, E and F. DCM's differ one from another by the types of SQUIDs they contain, which can be any two from D, E or F. In total fifteen DCM's were measured, six using SQUID type D, five type E and four type F. The DCM's are all made from YBCO films 220\,nm thick and were measured at 77\,K, so for inductance extraction we used $\lambda=402\,$nm, as in Table~\ref{tab:L_values}.

To simulate measurement of the effective area \Aeffw of the washer we added an extra conductor to the \mlsi model, designed to generate a reasonably uniform field \Bz through the \xyplane of the entire device.\cite{gerra:msthesis} If a current $I$ in this induces a flux $\Phi$ in the washer then
\begin{equation}\label{eqn:aeff1}
 \Aeffw = \Phi/\Bz = (\Mf I)/\Bz= \Mf/\alpha
\end{equation}
where \Mf is the mutual inductance between the field generator and washer, as simulated by \mlsi, and $\Bz=\alpha I$, with $\alpha$ found by Biot-Savart. We used a pair of in-plane parallel tracks, far away from either side of the device, which generate an acceptably-uniform \Bz. The same principle was used to find \Aeffsq for each of the three SQUID types used in the magnetometers.

Owing to the very different scales of the washer and the SQUID (as Fig.~\ref{fig:mw} shows) their \mlsi models need different FEM meshing arrangements and so we ran models for the washer and the SQUID separately. We found the washer self-inductance $\Lw=5,342\,$pH and \Aeffw=21.29\,mm$^2$. The SQUIDs had \Aeffsq in the range 398 to 720\,$\mu$m$^2$. We also extracted values of the coupled inductance \LCO for each SQUID type. The expected effective area of the magnetometer can then be calculated from the simulation data:
\begin{equation}
 \Aeffm = (\LCO/\Lw)\Aeffw + \Aeffsq  \approx (\LCO/\Lw)\Aeffw
\end{equation}
\Aeffsq is negligible, compared to \Aeffm and we also neglect mutual inductance between the washer and SQUID.

Experimentally  \Aeffm was found by applying a known magnetic field with a calibrated long solenoid and measuring over a minimum of $10\Phi_0$ with the flux-locked loop (FLL) unlocked. All measurements were made inside three layers of mu-metal shielding using Magnicon SEL FLL electronics.\cite{magnicon}

Table~\ref{tab:sq_results} summarizes the results of our simulations and the experimental values of \Aeffm. The measured \Aeffm are averages for each set of DCM's operating with the same type of SQUID.  We have also included simulated values of \Lsq, the total SQUID inductance, which are considerably higher than \LCO, due to the extra structure across the step-edge junctions with $2\,\mu$m wide tracks that closes the SQUID loop.

\begin{table}[h!]
\caption{\label{tab:sq_results}Data and effective areas for the three types of DCM. All have the same washer dimensions, but use SQUIDs that differ in the values of \lone. All have uncoupled lengths $\lzero=8\,\mu{\rm m}$. In this Table inductances are in pH, lengths in $\mu{\rm m}$ and areas in mm$^2$. The error is the fractional difference between measured and simulated data.}
\centering
\begin{center}
\begin{ruledtabular}
\begin{tabular}{cccccccccc}
SQUID  & $\ell_1$ & $s$ & $w_1$  & \LCO  &\Lw/\LCO & \Lsq & \Aeffm  & \Aeffm  & Error \\
type &&&&&&&(model)&(meas.)& \%\\
\hline
D    & 76 & 2 & 8 &  67.35 & 79.3 & 85.62 & 0.268 & 0.254 & 5.5\\
E    & 61 & 2 & 4 &  71.59 & 74.6 & 89.07 & 0.285 & 0.272 & 4.8\\
F    & 72 & 4 & 4 &  94.31 & 56.6 & 117.2  & 0.376 & 0.350 & 7.4\\

\end{tabular}
\end{ruledtabular}
\end{center}
\end{table}

In summary, by direct inductance measurements on twenty SQUIDs and inductance extraction by \mlsi we find $\lambda(77)$ is in the range 384 to 402\,nm for 220\,nm and 133\,nm films, with a mean of 391\,nm. We find $\lambda(77)$ is broadly independent of film thickness and film batch, which we believe confirms the high quality of all our films.  There are insufficient data to identify any clear dependence on \Tc. In general there is good agreement for each SQUID set between $\lambda_{\rm mlsi}$, $\lambda_{\rm ind}$ and $\lambda_{\rm fh}$, the values obtained at 77\,K using \mlsi, \induct and \FH. From SQUIDs with 220\,nm thick films measured in a cryo-cooler we found a close fit to (\ref{eq:lambda_T}) for $\lambda(T)$ between 50 and 79\,K. We view (\ref{eq:lambda_T}) as an empirical design aid for calculating the inductance of SQUID and SQIF loops to give optimum performance at any temperature. We used our inductance extraction methods to predict the effective areas of a directly-coupled magnetometers, which agree with the experimental value to better than 7.4\%.

The data that support the findings of this study are available from the corresponding authors upon reasonable request.

\bibliographystyle{style21} 
\raggedright

\bibliography{subset} 

\begin{thebibliography}{10}

\bibitem{keenan:298}
S.~T. Keenan, J.~Du, E.~E. Mitchell, S.~K.~H. Lam, J.~C. Macfarlane, C.~J.
  Lewis, K.~E. Leslie and C.~P. Foley, IEICE Trans. Electron. {\bf E96-C}, 298
  (2013).

\bibitem{Fagaly_Wiley_EncEEE}
R.~L. Fagaly, {\em {Superconducting Quantum Interference Devices (SQUIDs)}\/},
  pp. 1--15, John Wiley \& Sons, Inc. (2016).

\bibitem{stolz:033001}
R.~Stolz, M.~Schmelz, V.~Zakosarenko, C.~P. Foley, K.~Tanabe, X.~Xie and
  R.~Fagaly, Supercond. Sci. Technol. {\bf 34}, 033001 (2021).

\bibitem{lee:468}
J.~B. Lee, D.~L. Dart, R.~J. Turner, M.~A. Downey, A.~Maddever, G.~Panjkovic,
  C.~P. Foley, K.~E. Leslie, R.~Binks, C.~Lewis and W.~Murray, Geophysics {\bf
  67}, 468 (2002).

\bibitem{leslie:70}
K.~E. Leslie, R.~A. Binks, S.~K.~H. Lam, P.~A. Sullivan, D.~L. Tilbrook, R.~G.
  Thorn and C.~P. Foley, The Leading Edge {\bf 27}, 70 (2008).

\bibitem{keenan:025029}
S.~T. Keenan, J.~A. Young, C.~P. Foley and J.~Du, Supercond. Sci. Technol. {\bf
  23}, 025029 (2010).

\bibitem{linzen:71}
S.~Linzen, V.~Schultze, A.~Chwala, T.~Sch{\"u}ler, M.~Schulz, R.~Stolz and
  H.-G. Meyer, {\em Quantum Detection Meets Archaeology -- Magnetic Prospection
  with {SQUIDs}, Highly Sensitive and Fast\/}, pp. 71--85, Springer Berlin
  Heidelberg, Berlin, Heidelberg (2009).

\bibitem{NDE_SQUID_HB}
H.-J. Krause, M.~M\"uck and S.~Tanaka, in {\em Applied Superconductivity:
  Handbook on Devices and Applications\/} (edited by P.~Seidel), pp. 977--992,
  Wiley-VCH Verlag GmbH: Weinheim, Germany (2015).

\bibitem{mitchell:06LT01}
E.~E. Mitchell, K.~E. Hannam, J.~Lazar, K.~E. Leslie, C.~J. Lewis, A.~Grancea,
  S.~T. Keenan, S.~K.~H. Lam and C.~P. Foley, Supercond. Sci. Technol. {\bf
  29}, 06LT01 (2016).

\bibitem{oppenlander:936}
J.~Oppenl\"ander, C.~H\"aussler, A.~Friesch, J.~Tomes, P.~Caputo, T.~Tr\"auble
  and N.~Schopohl, IEEE Trans. Appl. Supercond. {\bf 15}, 936 (2005).

\bibitem{lam:123905}
S.~K.~H. Lam, R.~Cantor, J.~Lazar, K.~E. Leslie, J.~Du, S.~T. Keenan and C.~P.
  Foley, J. Appl. Phys. {\bf 113}, 123905 (2013).

\bibitem{fiory:2165}
A.~T. Fiory, A.~F. Hebard, P.~M. Mankiewich and R.~E. Howard, Appl. Phys. Lett.
  {\bf 52}, 2165 (1988).

\bibitem{claassen:3028}
J.~H. Claassen, M.~L. Wilson, J.~M. Byers and S.~Adrian, J. Appl. Phys. {\bf
  82}, 3028 (1997).

\bibitem{wang:3865}
R.~F. Wang, S.~P. Zhao, G.~H. Chen and Q.~S. Yang, Appl. Phys. Lett. {\bf 75},
  3865 (1999).

\bibitem{he:113903}
X.~He, A.~Gozar, R.~Sundling and I.~Bozovic, Rev. Sci. Instrum. {\bf 87},
  113903 (2016).

\bibitem{prozorov:R41}
R.~Prozorov and R.~W. Giannetta, Supercond. Sci. Technol. {\bf 19}, R41 (2006).

\bibitem{forrester:1835}
M.~G. Forrester, A.~Davidson, J.~Talvacchio, J.~R. Gavaler and J.~X. Przybysz,
  Appl. Phys. Lett. {\bf 65}, 1835 (1994).

\bibitem{grundler:5273}
D.~Grundler, B.~David and O.~Doessel, J. Appl. Phys. {\bf 77}, 5273 (1995).

\bibitem{fuke:L1582}
H.~Fuke, K.~Saitoh, T.~Utagawa and Y.~Enomoto, Jpn. J. Appl. Phys. {\bf 35},
  L1582 (1996).

\bibitem{mitchell:1282}
E.~E. Mitchell, D.~L. Tilbrook, C.~P. Foley and J.~C. MacFarlane, Appl. Phys.
  Lett. {\bf 81}, 1282 (2002).

\bibitem{li:1600404}
H.~{Li}, E.~Y. {Cho}, H.~{Cai}, Y.~{Wang}, S.~J. {McCoy} and S.~A. {Cybart},
  IEEE Trans. Appl. Supercond. {\bf 29}, 1600404 (2019).

\bibitem{ruffieux:025007}
S.~Ruffieux, A.~Kalaboukhov, M.~Xie, M.~Chukharkin, C.~Pfeiffer, S.~Sepehri,
  J.~F. Schneiderman and D.~Winkler, Supercond. Sci. Technol. {\bf 33}, 025007
  (2020).

\bibitem{ruffieux:asc2020}
S.~Ruffieux, N.~Lindvall, A.~Kalaboukhov, J.~F. Schneiderman and D.~Winkler,
  {\em Optimization of single layer high-${T}_{\rm c}$ {SQUID} magnetometers
  for low noise\/}, {ASC 2020 (unpublished)}.

\bibitem{shimazu:1451}
Y.~Shimazu and T.~Yokoyama, Physica C {\bf 412-14}, 1451 (2004).

\bibitem{lee:2419}
J.~Y. Lee and T.~R. Lemberger, Appl. Phys. Lett. {\bf 62}, 2419 (1993).

\bibitem{yoshida:3844}
K.~Yoshida, M.~S. Hossain, T.~Kisu, K.~Enpuku and K.~Yamafuji, Jpn. J. Appl.
  Phys. {\bf 31, Part 1}, 3844 (1992).

\bibitem{khapaev:1090}
M.~M. Khapaev, A.~Y. Kidiyarova-Shevchenko, P.~Magnelind and M.~Y. Kupriyanov,
  IEEE Trans. Appl. Supercond. {\bf 11}, 1090 (2001).

\bibitem{khapaev:055013}
M.~M. Khapaev and M.~Y. Kupriyanov, Supercond. Sci. Technol. {\bf 28}, 055013
  (2015).

\bibitem{foley:4281}
C.~Foley, E.~Mitchell, S.~Lam, B.~Sankrithyan, Y.~Wilson, D.~Tilbrook and
  S.~Morris, IEEE Trans. Appl. Supercond. {\bf 9}, 4281 (1999).

\bibitem{mitchell:065007}
E.~E. Mitchell and C.~P. Foley, Supercond. Sci. Technol. {\bf 23}, 065007
  (2010).

\bibitem{ceraco}
Ceraco ceramic coating GmbH, Ismaning, Germany http://ceraco.de.

\bibitem{star}
STAR Cryoelectronics, Santa Fe, USA http://starcryo.com.

\bibitem{kamon:1750}
M.~Kamon, M.~J. Tsuk and J.~K. White, IEEE Trans. Microw. Theory Techn. {\bf
  42}, 1750 (1994).

\bibitem{induct}
Available from Whiteley Research Inc., http://www.wrcad.com.

\bibitem{chang:764}
W.~H. Chang, IEEE Trans. Magn. {\bf 17}, 764 (1981).

\bibitem{ilichev:708}
E.~{Il'ichev}, L.~D\"orrer, F.~Schmidl, V.~Zakosarenko, P.~Seidel and
  G.~Hildebrandt, Appl. Phys. Lett. {\bf 68}, 708 (1996).

\bibitem{chen:9}
D.-X. Chen, C.~Navau, N.~Del-Valle and A.~Sanchez, Physica C {\bf 500}, 9
  (2014).

\bibitem{brorson:6185}
S.~D. Brorson, R.~Buhleier, J.~O. White, I.~E. Trofimov, H.-U. Habermeier and
  J.~Kuhl, Phys. Rev. B {\bf 49}, 6185 (1994).

\bibitem{prohammer:5370}
M.~Prohammer and J.~P. Carbotte, Phys. Rev. B {\bf 43}, 5370 (1991).

\bibitem{zaitsev:335}
A.~G. Zaitsev, R.~Schneider, G.~Linker, F.~Ratzel, R.~Smithey, P.~Schweiss,
  J.~Geerk, R.~Schwab and R.~Heidinger, Rev. Sci. Instrum. {\bf 73}, 335
  (2002).

\bibitem{farber:515}
E.~Farber, S.~Djordjevic, N.~Bontemps, O.~Durand, J.~Contour and G.~Deutscher,
  J. Low Temp. Phys. {\bf 117}, 515 (1999).

\bibitem{gerra:msthesis}
G.~Gerra, {\em Electromagnetic Modelling of Superconducting Sensor Designs\/},
  Master's thesis, University of Cambridge (2003).

\bibitem{magnicon}
Magnicon GmbH, Hamburg, Germany http://magnicon.com.

\end{thebibliography}

\end{document}